\documentclass{ws-p8-50x6-00}
\usepackage{amsfonts}
\def\P{{\mathbb P}}
\def\R{{\mathbb R}}
\def\J{{\mathbb J}}
\def\W{{\mathbb W}}
\def\F{{\mathbb F}}
\def\K{{\mathbb K}}
\def\T{{\mathbb T}}

\begin{document}

\title{On Scission Configuration in ternary fission}

\author{V. G. Kartavenko$~^{1,2}$, A. S\u{a}ndulescu$^{2,3}$
and W. Greiner$~^2$}

\address{
$^1$~Bogoliubov Laboratory of Theoretical Physics,
Joint Institute for Nuclear Research,
Dubna, Moscow District, 141980, Russia\\[1mm]
$^2$~Institut f\"ur Theoretische Physik der J. W. Goethe Universit\"at\\
D-60054 Frankfurt am Main, Germany\\[1mm]
$^3$~Romanian Academy, Calea Victoriei 125, Bucharest, 71102, Romania}


\maketitle

\abstracts{
A static scission configuration in cold ternary fission
has been considered in the framework of two mean field approaches.
The virial theorems has been suggested to investigate correlations in
the phase space, starting from a kinetic equation.
The inverse mean field method is applied
to solve single-particle Schr\"odinger equation,
instead of constrained
selfconsistent Hartree--Fock equations.
It is shown,
that it  is possible to simulate one-dimensional three-center system
via inverse scattering method
in the approximation of reflectless single-particle potentials.}

\section{Motivation}
Ternary fission involving the emission of
$\alpha$-particle was first observed
more than fifty years ago \cite{alvarez}.
Emission of $\alpha$-particles in the spontaneous fission
of $^{252}$Cf has also a long history of investigation
experimentally \cite{fraenkel67} and  theoretically
\cite{fraenkel67etc,sandulIJMPE99}.
A renewed interest in these processes
arose in connection with
\begin{itemize}
\item{}
modern experimental techniques
($\gamma-\gamma-\gamma$ and $x-\gamma-\gamma$
triple coincidence,
(Gammasphere with 110 Compton suppressed
Ge detectors), which allow
the fine resolution of the mass, charge
and angular momentum content of the fragments.)\cite{terakopian94}
\item{}
Recently direct experimental evidence was presented
for the cold  (neutronless) ternary spontaneous fission of $^{252}$Cf
in which the third particle is an $\alpha$-particle
\cite{He4ternary}, or $^{10}$Be \cite{Be10ternary}.
This confirms that a large variety of nuclear
large-amplitude collective motions such as bimodal fission \cite{hulet86},
cold binary  fission
\cite{terakopian94,Hamb93,Schwab94,Dard96,sandul96},
heavy cluster radioactivity \cite{spg80,price89},
and inverse processes, such as subbarier fusion \cite{armbruster}, could
belong to the general phenomenon of cold nuclear fragmentation.
\item{}
Cluster like models \cite{sandul89,goenborsig90}
were used successfully to reproduce general features of
the cold ternary fragmentation. However the scission configuration has been
built in fact by hands.
\end{itemize}

Therefore it is actual to develop microscopical or
semi-microscopical approach to this scission-point concept of nuclear
fragmentation.
There are well developed methods to calculate, in the framework of
many-body self--consistent approach,
static properties
of a well isolated nucleus in its ground state.
There also exists a well developed two-center shell model \cite{2csh72}.
However, a three-center shell model has not been developed yet,
except for very early steps \cite{sg3c1977}.
Three-center shapes are practically not
investigated, in comparison with the two-center ones.
There exists the  generalizations of mean--field models to the case of
two-centers \cite{berger89}, but
a ternary configuration is out of consideration,
because of uncertainties to select
a peculiar set of constraints.

There exist a number of calculations for nucleus-nucleus collisions
in the frame of time-dependent mean--field methods, but an evolution of
the cold fragmentation has not been investigated yet.
Therefore, although the principal way to describe nuclear fragmentation
in the framework of many-body self-consistent approach exists,
it is interesting to
develop other mean-field approaches to analyse these phenomena
from different points of view.

In this Contribution we suggest two methods to analyse
a static scission configuration in cold ternary fission
in the framework of mean field approach.
In Section 2, starting from kinetic equation
the virial theorems has been suggested to investigate correlations in
the phase space. It gives possibility to formulate in future
generalised set
of constraints in momentum and cartesian spaces.
In Section 3
the inverse mean field method is applied
to solve single-particle Schr\"odinger equation,
instead of constrained
selfconsistent Hartree--Fock equations.
It makes irt possible to simulate one-dimensional three-center system
via inverse scattering method.

\section{Virial theorems}
Within the mean-field approximation, we analyze
the evolution of one-body
Wigner phase-space distribution function of the full
many-body wave function,
following the well developed
scheme using  as the starting point
the Vlasov equation
for the Wigner phase-space distribution function \cite{wigner}
\begin{equation}
\frac{\partial f}{\partial t}+\frac{\vec{p}}{m}\cdot
\frac{\partial f}{\partial{\vec r}}-\frac{\partial V}{\partial
{\vec r}}\cdot
\frac{\partial f}{\partial\vec{p}}=I_{rel},
\label{vlasov}
\end{equation}
with a "relaxation term" $I_{rel}$ is added to the kinetic equation
to describe dissipation effects.
The quantity $V(\vec{r},t)$ is the self--consistent single-particle
potential which is assumed here to be local,
$m$ is the mass of nucleon.\\
Out of the Wigner distribution function,
virials at different orders can help to extract
useful physical information from the total phase space dynamics.
Integrating the initial kinetic equation (\ref{vlasov})
over the momentum space with different
polynomial weighting functions of the ${\vec p}$-variable
one comes, as well know~\cite{chandr},\cite{rozensteel},\cite{balb91},
to an infinite chain of equations
for local collective observables including the density, collective
velocity,
pressure and an infinite set of tensorial functions of the time and space
coordinates, which are defined as moments of the distribution function in
the momentum space:
\begin{itemize}
\item
the particle $n({\vec r},t)\equiv \int d\vec{p}\,f({\vec r},\vec{p},t),$
and the mass $\rho({\vec r},t)=m\,n({\vec r},t)$ densities,
\item
the collective current and velocity of nuclear matter
\begin{eqnarray}\nonumber
\rho ({\vec r},t){\vec  u}({\vec r},t)=
\int d\vec{p}\,\vec{p}f({\vec r},\vec{p},t),
\end{eqnarray}
\item
the pressure tensor and the energy and momentum transfer tensors
of different orders
\begin{eqnarray}\nonumber
\P_{ij}({\vec r},t)&=&\frac{1}{m}\int d\vec{p}\, q_{i}
q_{j}f({\vec r},\vec{p},t),\qquad
q_i\equiv p_i-mu_i,\\ \nonumber
\P_{\underbrace{ij..k}_{n}}({\vec r},t)&=&
\frac{1}{m^{n-1}}\int d\vec{p}\,
\underbrace{q_{i}q_{j}..q_{k}}_{n} f({\vec r},\vec{p},t),
\end{eqnarray}
\item
and the  integrals related to relaxation terms
\begin{eqnarray}\nonumber
\int d\vec{p}\,I_{rel}&=&0,\qquad
\int d\vec{p}\, \vec{p} I_{rel}=0,\\ \nonumber
\R_{ij} &\equiv& \frac{1}{m}
\int d\vec{p}\,q_{i}q_{j}I_{rel},\qquad\dots
\end{eqnarray}
\end{itemize}
Truncating  this chain at order two in $\vec{q}$
one arrives at the "fluid
dynamical" level of description of nuclear processes.
\begin{equation}
\frac{\partial \rho}{\partial t} + \sum\limits_{k}
\frac{\partial}{\partial x_{k}} (u_{k}\rho) = 0,
\label{hyd1}
\end{equation}
\begin{eqnarray}\nonumber
\rho \frac{D u_{i}}{D t}  &+& \sum\limits_{k}
\frac{\partial \P_{ik}}{\partial x_{k}}
+ \frac{\rho}{m} \frac{\partial V}{\partial x_{i}}
+ \rho (\Omega_{i} \sum\limits_{k}\Omega_{k}x_{k} - \Omega^{2} x_{i} )\\
&+& \rho \sum\limits_{s,j} \varepsilon_{isj} (2\Omega_{s} u_{j} +
\frac{d\Omega_{s}}{d t}x_{j}) = 0,
\label{hyd2}
\end{eqnarray}
\begin{eqnarray}\nonumber
\frac{D \P_{ij}}{D t} &+& \sum_{k}\left( \P_{ik}
\frac{\partial u_{j}}{\partial x_{k}} + \P_{jk}
\frac{\partial u_{i}}{\partial x_{k}} +
\P_{ij}\frac{\partial u_{k}}{\partial x_{k}}\right) \\ \nonumber
&+& 2\sum\limits_{s,m} \Omega_{m} (\varepsilon_{jms} \P_{is} +
\varepsilon_{ims}\P_{js})\\
&+&\sum\limits_{k}\frac{\partial}{\partial x_{k}}\P_{ijk} =
\left( \frac{\partial \P_{ij}}{\partial t}\right)_{rel},\\ \nonumber
\left(\frac{\partial \P_{ij}}{\partial t}\right)_{rel} &\equiv&
\frac{1}{m}\int d\vec{p}\;q_i q_j I_{rel},
\label{hyd3}
\end{eqnarray}
where the usual notation
$\frac{D}{D t}\equiv \frac{\partial }{\partial t}
+ \sum\limits_{k} u_{k} \frac{\partial}{\partial x_{k}}$
is introduced for the operator giving the material derivative,
or the rate of change at a point moving locally with the fluid.
The hydrodynamical set of Eqs.~(\ref{hyd1}-\ref{hyd3})
describes an evolution of a rotating nuclear system.
We consider two frames of reference with a common origin:
an inertial frame, ($X_{1}, X_{2}, X_{3}$), and a moving frame,
($x_{1}, x_{2}, x_{3}$).
Let $x_{i} = \sum\limits_{j=1}^{3} \T_{ij} X_{j}$
be the linear transformation that relates the coordinates,
$\vec{X}$ and $\vec{x}$, of a point
in two frames. The orientation of the moving frame,
with respect to the inertial frame, will be assumed to be time dependent.
Since $\T_{ij}(t)$ must represent an orthogonal transformation,
the vector
\begin{eqnarray}\nonumber
\Omega_{i} = \frac{1}{2}\sum_{j,k,m}\varepsilon_{ijk}
\left(\frac{d\T}{d t}\right)_{jm} \T^{+}_{mk},
\label{omega}
\end{eqnarray}
represents a general time-dependent rotation
of the $\vec{x}$-frame with respect to the inertial frame.\\
Let us define integral collective "observables"
(the integrals over the whole phase space of one nucleon containing the
distribution function appropriatly weighted), namely
an inertia tensor $\J_{ij}(t)$,
the dynamical part of the angular momentum  $L_{i}(t)$,
the integral pressure tensor
${\Pi}_{ij}(t)$ defined as
\begin{eqnarray}\nonumber
\J_{ij}&\equiv& \int d\vec{x}\;x_{i}x_{j}\rho,\qquad
{\Pi}_{ij}\equiv\int d\vec{x}\;\P_{ij},\\ \nonumber
L_{k} &\equiv& \sum_{i,j} \varepsilon_{kij} \int d\vec{x}\;
\rho x_{i}u_{j},
\end{eqnarray}
The dynamics in terms of the latter "observables" is expressed by a set of
virial equations in the rotating frame~\cite{kmmq2k}
\begin{eqnarray}\nonumber
\frac{d^{2}}{d t^{2}} \J_{ij}
&+&
\sum\limits_{k} \Omega_{k} (\Omega_{i} \J_{jk}+\Omega_{j} \J_{ik})
-2\Omega^{2} \J_{ij}\\ \nonumber
&+& 2 \sum\limits_{s,k} \Omega_{s}
\int d{\vec r}\; \rho u_{k}
(\varepsilon_{isk} x_{j} + \varepsilon_{jsk} x_{i})\\ \nonumber
&+& 2\W_{ij} - 2\K_{ij} -2\Pi_{ij}\\ \nonumber
&+&
 \sum\limits_{s,k} \frac{d \Omega_{s}}{d t}
(\varepsilon_{isk} \J_{kj}+\varepsilon_{jsk} \J_{ki}) = 0,\\
\nonumber
\frac{d L_{k}}{d t} &+& \sum_{i,j,m}
\varepsilon_{kji}\Omega_{i}\Omega_{m}\J_{jm} -
2\sum_{s}\Omega_{s}\int d{\vec r}\; \rho u_{k} x_{s} \\  \nonumber
&-& \sum_{s} \frac{d\Omega_{s}}{d t} \J_{ks} + \frac{d}{d t}
(\Omega _{k} \sum_{j} \J_{jj}) = 0,\\ \nonumber
\frac{d}{d t} {\Pi}_{ij} &+& \F_{ij}
+ 2 \sum\limits_{s,k} \Omega_{s} (\varepsilon_{isk} {\Pi}_{kj} +
\varepsilon_{jsk} {\Pi}_{ki} ) = \R_{ij},\\ \nonumber
\F_{ij} &\equiv&
\sum\limits_{k} \int d{\vec r}\; \left(
\P_{ik}\frac{\partial u_{j}}{\partial x_{k}} +
\P_{jk}\frac{\partial u_{i}}{\partial x_{k}} \right),
\label{vir5}
\end{eqnarray}
where the tensors of collective kinetic and potential energies,
and the relaxation tensor are
\begin{eqnarray} \nonumber
\K_{ij}&=& \int d\vec{x}\;u_{i}u_{j} \rho, \qquad
\W_{ij}=\int d\vec{x}\;x_{j}\frac{\partial V}{\partial x_{i}}n,\\
\nonumber
\R_{ij}&\equiv&\int d\vec{x}\;
\left(\frac{\partial \P_{ij}}{\partial t}\right)_{rel}
x_{i} x_{j}.
\end{eqnarray}
The above equations constitute
a formal framework within which the
coupling of the deformations in the ${\vec r}$-space
and in the ${\vec p}$-space can be explicitly worked out.
It may help to fofmulate in a future a generalised set of
constrains in the phase space.
%
\section{Inverse mean field method}
\subsection{The framework}
%
Methods of nonlinear dynamics.
gave yet the possibility to derive
for nuclear physics unexpected collective modes,
which can not be obtained by traditional  methods of perturbation
theory near some equilibrium state (see e.g. review
\cite{kart93pn} and \cite{KSG99} for the recent refs.).
The most important reason is that
the fragmentation and clusterization is a very general phenomenon.
There are cluster objects
in subnuclear and macro physics.
Very different
theoretical methods were developed in these fields. However,
there are only few basic physical ideas, and most of the methods
deal with nonlinear partial differential equations.
One of the most important part of soliton theory is the inverse scattering
method \cite{gelfand51,marchenko55,faddev59}
and its applications to the integration of nonlinear partial
differential equations \cite{novikov70}.
The inverse methods to integrate nonlinear evolution equations
are often more effective than a direct numerical integration.
Let us demonstrate this statement for the following simple case.
The type of systems under consideration are slabs of
nuclear matter \cite{bcn76}, which are finite in the $z$ coordinate and
infinite and homogeneous in two transverse directions.
The wave function for the slab geomethry is
\begin{equation}
\psi_{{{\mathbf k}_{\perp}} n}
({\mathbf x}) = {1 \over {\sqrt{\Omega}}} {\psi}_{n} (z)
\exp (i {\mathbf k}_{\perp}{\mathbf r}),\qquad
\epsilon_{{{\mathbf k}_{\perp}} n} = {\hbar^{2} k_{\perp} ^{2}
\over 2m} + e_{n},
\label{eq:slab}
\end{equation}
where ${\mathbf r} \equiv (x,y), \/ {{\mathbf k}_{\perp}}
\equiv (k_{x}, k_{y})$, and $\Omega$ is the transverse normalization area.\\
\begin{equation}
-{\hbar^{2} \over 2m} {d^{2} \over dz^{2}}
\psi_{n} (z) + U(z) \psi_{n} (z) = e_{n} \psi _{n} (z),
\label{eq:sp-problem}
\end{equation}
A direct method to solve the single-particle problem  (\ref{eq:sp-problem})
is to assign a functional of interaction ${\cal E}$
(usually an effective density dependent Skyrme force),
to derive the ansatz for the one-body potential,
as the first variation of a functional of interaction in density
%
$U(z) = U[\rho(z)] = \delta{\cal E}/\delta\rho.$
%
Then to solve the Hartree-Fock problem under the set of constraints,
which define the specifics of the nuclear system.
In the simplest case of a ground state, one should conserve the total particle
number of nucleons ($A$), which is related to the "thickness" of a slab, via
$A \Longrightarrow {\cal A} = ( 6 A {\rho _{N} ^{2}} / \pi ) ^{1/3}$,
which gives the same radius for a three-dimensional system and
its one-dimensional analogue.
As a result, one obtains the energies of the single particle states
$e_{n}$, their wave functions $\psi _{n} (z)$, the density profile
$\rho ({\mathbf x}) \Longrightarrow \rho (z)$
\begin{equation}
\rho (z) =
\sum _{n=1} ^{N_{0}} a_{n} \psi _{n} ^{2} (z),\qquad
{\cal A} =
\sum _{n=1} ^{N_{0}} a_{n}, \qquad
a_{n} = {2m \over \pi \hbar^{2} } ( e_{F} - e_{n}),
\label{eq:density}
\end{equation}
and the corresponding single-particle potential.
$a_{n}$ are the occupation numbers,
$ N_{0}$ is the number of  occupied bound orbitals.
The Fermy-energy $e_{F}$ controls
the conservation of the total number of nucleons.
%
%
The energy (per nucleon) of a system  is given by
\begin{equation}
{E \over A} \Longrightarrow { \hbar^{2}
\over 2m{\cal A}} \Bigr(
 \sum _{n=1} ^{N_{0}} a _{n} \int _{-\infty} ^{\infty}
\bigr( {d \psi_{n} \over dz} \bigl) ^{2} dz
+ {\pi \over 2} \sum _{n=1} ^{N_{0}} a_{n} ^{2} \Bigl)
 + {1 \over {\cal A}} \int _{-\infty} ^{\infty}
{\cal E} [ \rho (z) ] dz.
\label{eq:energy}
\end{equation}
Finally, the set of formulas (\ref{eq:slab}--\ref{eq:energy}) completely
defines the direct self-consistent problem.
Following the inverse scattering method,
one reduces the main  differential Schr\"odinger equation
(\ref{eq:sp-problem}) to
the integral Gel'fand-Levitan-Marchenko equation
\cite{gelfand51,marchenko55}
\begin{equation}
K (x, y) + B (x + y) + \int _{x} ^{\infty} B (y + z) K (x,z) dz =0.
\label{eq:GLM}
\end{equation}
for a function $K(x,y)$.
The kernel $B$ is determined by the reflection coefficients
$R (k) \/ ( e_{k} = \hbar^{2} k^{2} / 2m )$,
and by the $N$ bound state eigenvalues
\[
B (z) = \sum _{n=1} ^{N} C_{n} ^{2} ( \kappa_{n} )  +
{1 \over \pi} \int _{-\infty} ^{\infty} R (k) \exp \/ (ikz)\/ dk, \qquad
e_{n} = - \hbar^{2} \kappa_{n} ^{2} / 2m .
\]
$N $ is the total number of the bound orbitals.
The coefficients $C _{n}$ are uniquely specified by the boundary conditions
and the symmetry of the problem under consideration.
The general solution,
%
$U (z) = - (\hbar^{2} / m) (\partial K (z, z)/ \partial z),$
%
should naturally contain both,
contributions due to the continuum of the spectrum and to its discrete
part. There seems to be no way to obtain
the general solution $U(z)$ in a closed
form. Eqs.~(\ref{eq:sp-problem}),(\ref{eq:GLM})
have to be solved only numerically.
In Ref. \cite{kartmad87} we used the approximation of
reflectless ($R(k) = 0$),
symmetrical ($U(-z) = U(z)$) potentials.
This gave the possibility to derive
the following set of relations
\begin{eqnarray} \nonumber
U (z) = - {\hbar^{2} \over m} {\partial^{2} \over \partial z^{2}}
\ln ( \det \Vert M \Vert ) = - {2 \hbar^{2} \over m}
\sum _{n=1} ^{N} \kappa_{n} \psi _{n} ^{2} (z),\\ \nonumber
\psi _{n} (z) = \sum _{n=1} ^{N} ( M^{-1} )_{nl} \lambda _{l} (z),\qquad
\lambda _{n} (z) = C_{n} (\kappa_{n}) \exp \/ ( - \kappa_{n}z ),\\
M_{nl} (z) = \delta_{nl} + {{\lambda_{n} (z) \lambda_{l} (z)} \over
{\kappa_{n} + \kappa_{l}}},\qquad
C_{n} (\kappa_{n}) = \Bigl( 2 \kappa_{n}
\big{\vert} {\prod _{l {\not=} n} ^{N}}
{{\kappa_{n} + \kappa_{l}} \over {\kappa_{n} - \kappa_{l}}}
\vert \Bigr) ^{1/2}.
\label{eq:inverse}
\end{eqnarray}
Consequently, in the approximation of reflectless potentials ($R(k) = 0$),
the wave functions, the single-particle potential and the density profiles
are completely
defined by the bound state eigenvalues via formulas
(\ref{eq:slab}),(\ref{eq:inverse}).
%
%
\subsection{Results and Discussion}
In Ref. \cite{kartmad87} a series of calculations for the
different layers, imitating nuclear
systems in their ground state was provided.
For a direct part of the calculations by the Hartree-Fock method, the
interaction functional was chosen in the form of effective Skyrme forces.
The calculated spectrum of bound states was fed into the scheme of the inverse
scattering method, and the relations
were used to recover the wave functions of
the states, the single-particle  potentials, and the corresponding densities.
In this note, we generalize this method to the case of
fragmented  configuration, trying to imitate two- and three-center
nuclear systems. We use here only the inverse mean-field scheme
(\ref{eq:inverse}). The details of the approach and
systematic calculations of fragmented nuclear systems will be
provided in a forthcoming publication.
\begin{figure}[t]
\epsfxsize=79mm 
\figurebox{120mm}{80mm}{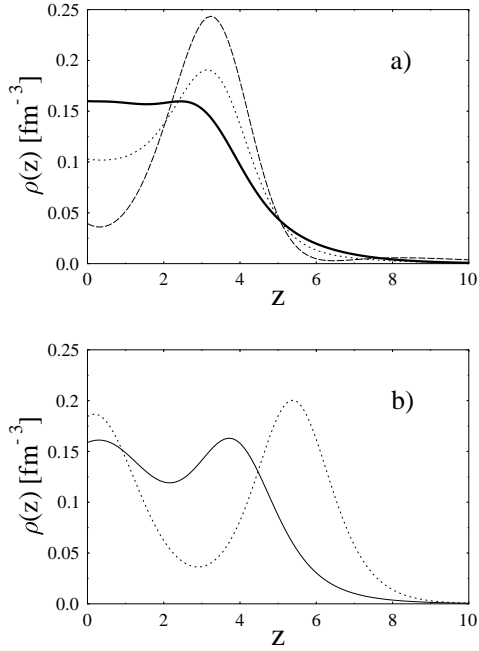} 
\caption{The density profiles of the light
($A\approx 20,\; {\cal A}\approx~1.0$)
three-levels ($N=3,\; N_0=3$)
model system calculated in the frame of inverse mean field method.
a) the ground state (solid line); a fragmented two-center configuration
(dotted and dashed lines);
b) a ternary fragmentation of the system into three fragments.
}
\label{fig}
\end{figure}
%
In Fig.1 we present the results of the calculations of three-level
($N=3,\; N_0=3$) light ($A\approx 20,\; {\cal A}\approx~1.0$)
model system simulating the ground state
(Fig.~\ref{fig}(a) solid line), and
fragmentation of the system into two fragments (Fig.~\ref{fig}(a),
the dotted and dashed lines).
In the same figure (Fig.~\ref{fig}(b)) we  present the
fragmentation of the system into three fragments (solid and dotted
lines).
One can see that it is possible to simulate
a one-dimensional three-center system
via inverse scattering method.
The following conclusions can be drawn.
\begin{itemize}
\item{}
The density profiles,
calculated in the framework of inverse method, are practically
identical to the results of calculation by SHF
method. These results are valid for the ground state and
for the system in the external potential field.
\item{}
The global properties of single-particle potentials
(the depth and an effective radius)
have been reproduced quite well, but the
inverse method yields
the quite strongly pronounced oscillations of the potential
distributions within the inner region, and
slightly different asymptotic tails of potential.
In the framework of inverse scattering method, all bound states are taken into
account in the calculation of the potential (\ref{eq:inverse}), but for the
density distribution only the occupied states are taken into account
(see Eqs.~(\ref{eq:density})).
Therefore, the slope of the tails of the potential and of
the density distributions will we different.
\item{}
We used, the approximation of reflectionless potentials, which gave us the
possibility to obtain a simple closed set of relations (\ref{eq:inverse}),
to calculate wave functions, density distributions and single particle
potentials.
The omitted reflection terms ($R(k)=0$)are not important
for the evaluation of the density
distributions, due to the fact that only
the deepest occupied states are used to evaluate density distribution
(see Eq.~(\ref{eq:density})).
The introduction of these reflection terms will lead to a smoothing of the
oscillations in the inner part of the potential and to a correction of its
asymptotic behaviour.
\item{}
The presented method gives a tool to simulate the  various sets of the
static excited states of the system.
This method could be useful to
prepare in a simple way an initial state for the
dynamical calculations in the frame of mean-field methods.
\end{itemize}
\section{Conclusions}
Recent experimental progress in the investigation of cold nuclear
fragmentation has made the development of theoretical many-body methods
highly desirable.
Modern variants of self-consistent Hartree-Fock and
relativistic mean-field models give the principal way
to describe nuclear fragmentation in the
framework of many-body self-consistent approach.
However, the generalization of these approaches to three-center
case is not provided yet because of existing difficulties to select a
suitable set of constraints.

We suggest two methods to analyse
a static scission configuration in cold ternary fission
in the framework of mean field approach.
The virial theorems has been suggested to investigate correlations in
the phase space, starting from a kinetic equation.
The inverse mean field method is applied
to solve single-particle Schr\"odinger equation,
instead of constrained
selfconsistent Hartree--Fock equations.
It is shown,
that it  is possible to simulate one-dimensional three-center system
via inverse scattering method
in the approximation of reflectless single-particle potentials.

These models may be useful as a guide to understand
the general properties
of fragmented systems and to formulate the suitable set of constraints
for the realistic three-dimensional mean field calculations of the
three-center nuclear system.
%
%
\section*{Acknowledgments}
The partial financial support
by Russian Foundation for Basic Research and
Deutsche Forschungsgemeinschaft is gratefully acknowledged.
%
%

\end{document}